\documentclass[conference]{IEEEtran}
\IEEEoverridecommandlockouts
\usepackage{cite}
\usepackage{amsmath,amssymb,amsfonts}
\usepackage{algorithmic}
\usepackage{graphicx}
\usepackage{textcomp}
\usepackage{threeparttable}
\usepackage{multirow}
\usepackage{tabularx}
\usepackage{booktabs}
\usepackage{xcolor}
\usepackage{subfigure}
\def\BibTeX{{\rm B\kern-.05em{\sc i\kern-.025em b}\kern-.08em
    T\kern-.1667em\lower.7ex\hbox{E}\kern-.125emX}}
\begin{document}

\title{Guiding the underwater acoustic target recognition with interpretable contrastive learning\\
}

\author{\IEEEauthorblockN{1\textsuperscript{st} Yuan Xie 2\textsuperscript{nd} Jiawei Ren 3\textsuperscript{rd} Ji Xu}
\IEEEauthorblockA{\textit{Key laboratory of Speech Acoustics and Content Understanding, Institute of Acoustics, Chinese Academy of Sciences} \\
\textit{University of Chinese Academy of Sciences} Beijing, China \\
\{xieyuan, renjiawei, xuji\}@hccl.ioa.ac.cn}

}

\maketitle

\begin{abstract}
Recognizing underwater targets from acoustic signals is a challenging task owing to the intricate ocean environments and variable underwater channels. While deep learning-based systems have become the mainstream approach for underwater acoustic target recognition, they have faced criticism for their lack of interpretability and weak generalization performance in practical applications. In this work, we apply the class activation mapping (CAM) to generate visual explanations for the predictions of a spectrogram-based recognition system. CAM can help to understand the behavior of recognition models by highlighting the regions of the input features that contribute the most to the prediction. Our explorations reveal that recognition models tend to focus on the low-frequency line spectrum and high-frequency periodic modulation information of underwater signals. Based on the observation, we propose an interpretable contrastive learning (ICL) strategy that employs two encoders to learn from acoustic features with different emphases (line spectrum and modulation information). By imposing constraints between encoders, the proposed strategy can enhance the generalization performance of the recognition system. Our experiments demonstrate that the proposed contrastive learning approach can improve the recognition accuracy and bring significant improvements across various underwater databases.

% The visualized results can help understand the behavior of recognition models by analyzing the regions that contribute the most to the prediction, which is highlighted in the CAM heat map.

\end{abstract}


\section{Introduction}
Underwater acoustic target recognition is a component of ocean acoustics. It is a crucial task that aims to realize automatic recognition of the radiating sound of targets\cite{b1}. Developing a robust recognition system is essential for monitoring maritime traffic and identifying the source of noise in underwater environmental monitoring systems\cite{b2}.

%Early work in underwater acoustic recognition applied manually-design features as inputs, such as audio energy, zero-crossing rate, low-frequency analysis recording (LOFAR)\cite{b3} or detection of envelope modulation on noise (DEMON)\cite{b4}. However, low-dimensional manual features inherently limit the information comprehensiveness and generalization ability for data with diverse feature space\cite{b5}. With the development of deep learning and the accumulation of underwater acoustic database\cite{b5}\cite{b6}, recognition algorithm based on deep learning continues to grow in popularity. Considering that deep neural networks could effectively utilize high-dimensional nonlinear data by automated learning, many works take time-frequency spectrograms as the input of neural networks\cite{b7} and achieve good successes. According to the research status, deep learning-based paradigms tend to outperform traditional methods in underwater acoustic recognition.

With the development of deep learning and the accumulation of underwater acoustic databases\cite{b5}\cite{b6}, recognition systems based on deep neural networks gradually become mainstream. Compared with classic machine learning paradigms,  which may not achieve promising performance for data with diverse feature space\cite{b5}, deep learning-based methods can capture high-level information from acoustic features and achieve significant successes\cite{b3}\cite{b4}\cite{b7} across existing underwater databases. Although noticeable improvement has been achieved, the performance of deep learning-based recognition models is still not satisfactory in practical applications\cite{b1}. Furthermore, the complex topology and nonlinearities of deep neural networks limits their interpretability. A question is raised: ``Where and how do neural networks learn discriminative information from the acoustic features?''

\begin{figure}[t]
    \centering
    \includegraphics[width=\linewidth]{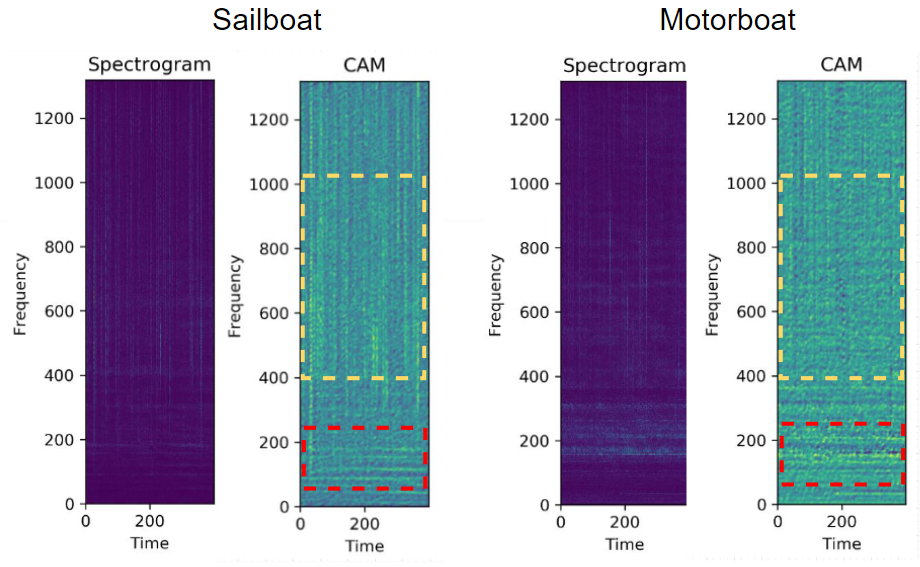}
    \caption{The class activation mapping (CAM) heat maps for two samples in Shipsear. The heat maps highlight the line spectrum (see red box) in the low-frequency component and the periodic modulation information (see yellow box) in the high-frequency component.}
    \label{fig1}
\vspace{-2mm}
\end{figure}

 \begin{figure*}[t]
    \centering
    \includegraphics[width=0.95\linewidth]{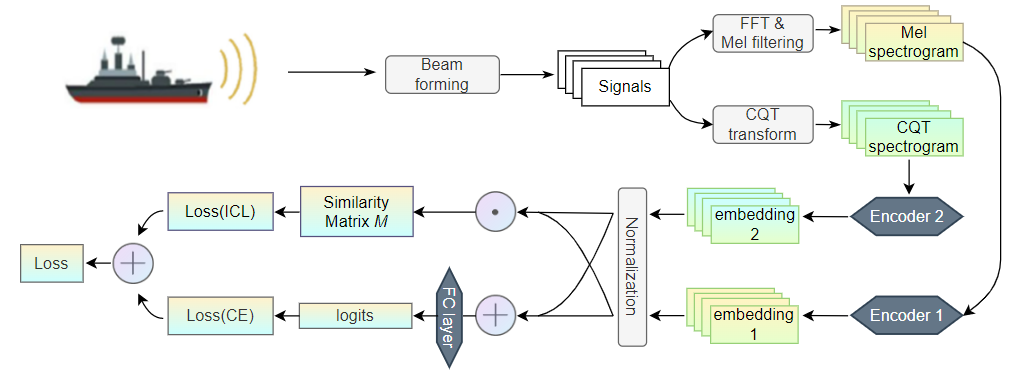}
    \caption{The preprocessing and training pipeline of our recognition system. Gray boxes represent operations and black boxes represent neural network modules. The detailed flow of feature extraction is omitted for simplicity.}
    \label{fig2}
\vspace{-2mm}
\end{figure*}

To address these issues, we employ the class activation mapping (CAM)\cite{b8} technique to highlight and visualize the critical regions in the spectrogram. CAM helps to understand the behavior of recognition models by highlighting the regions of the input features that contribute the most to the prediction. It generates a heat map of the input features, where the regions that contributed the most to the prediction are shown in bright colors and the less important regions are shown in dull colors. As shown in Fig.~\ref{fig1}, the CAMs on time-frequency spectrograms highlight the low-frequency line spectrum information and high-frequency periodic modulation information that guide the model learning.

Based on this observation, we propose a recognition system that simultaneously focuses on these two information. In this work, we apply two acoustic features with different emphases as the input features. To enhance the attention of recognition models to line spectrum and modulation information, we respectively extract the Mel spectrogram with higher frequency resolution at low frequencies and the CQT spectrogram with high temporal resolution at high frequencies. Then, we employ two encoders as the multi-perspective learner to learn from acoustic features with different emphases, and implement interpretable contrastive learning (ICL) as the training strategy to learn comprehensive and easy-to-generalize knowledge. Specifically, we apply the cosine-similarity-based contrastive learning strategy\cite{b9}\cite{b10} to simultaneously optimize two encoders. Our experiments show that ICL achieves superior performance compared to traditional spectrogram-based recognition techniques. Moreover, the contrastive learning can constrain the two encoders by encouraging the similarity between their outputs, thus achieving a regularization effect. Compared with strategies that fuse or ensemble two acoustic features, ICL can bring stronger robustness and generalization ability to recognition systems.

\section{Related Works}

\subsection{Acoustic Features and Recognition Algorithms}
In recent years, the increasing demand has fostered research aiming at optimizing acoustic features and recognition algorithms. Early work applied classic machine learning techniques to process low-dimensional acoustic features. Das et al.\cite{b11} applied cepstral features with the cepstral liftering process and the Gaussian mixture model (GMM) to classify marine vessels; Wang and Zeng\cite{b12} used a bark-wavelet analysis combined with the Hilbert–Huang transform to analyze signals, and employed support vector machines (SVMs) as the classifier. Moreover, acoustic features from the audio and speech domains (e.g., Mel frequency cepstrum coefficients) are also widely applied to ship-radiated noise recognition tasks and show promising results\cite{b13}\cite{b14}. However, some studies revealed that low-dimensional acoustic features and classical machine learning models are not satisfactory for large-scale data with diverse feature space\cite{b5}.

With the development of deep learning\cite{b15} and the accumulation of underwater acoustic databases\cite{b5}\cite{b6}, recognition algorithm based on deep neural networks continues to grow in popularity. As reported in the literature, Zhang et al.\cite{b16} applied the short-time Fourier transform (STFT) amplitude spectrum, STFT phase spectrum, and bispectrum features as the input for the convolutional neural network\cite{b17}; Liu et al.\cite{b18} used convolutional recurrent neural networks with 3-D Mel-spectrogram and data augmentation to conduct underwater target recognition; Xie et al.\cite{b7} applied learnable fine-grained wavelet transform and the deep residual network (ResNet)\cite{b19} to adaptively recognize ship-radiated noise; Ren et al.\cite{b23} utilized learnable Gabor filters and ResNet to build an intelligent underwater acoustic classification system... Compared with classic methods, deep nerual networks can utilize vast parameters and complex nonlinearities to learn high-level knowledge from high-dimensional features (e.g., STFT spectrogram, Mel spectrogram, etc.), which contain more comprehensive time–frequency information.

\subsection{Multi-Feature Learning}
The multi-feature learning is a technique used to fuse or ensemble different acoustic features to improve the performance on audio-related tasks. Considering that different acoustic features behave differently in complex ocean environments, multi-feature learning can take advantage of multiple acoustic features to complement each other, thereby improving the robustness of the model. Zhang et al.\cite{b16} integrated the STFT amplitude spectrum, phase spectrum, and bispectrum features as the model input; Cai et al.\cite{b20} dynamically fused multiscale significant features and spatial semantic features to recognize underwater weak targets; Hong et al.\cite{b21} combined the zero-cross ratio, log Mel spectrogram, MFCCs, chromatogram as aggregated features along with the data augment strategy... The consensus of many works is that the fusion of multiple features can usually achieve better recognition performance than single-feature-based methods. Besides, Xie et al.\cite{b1} integrated labeled ocean environment information (distance, channel depth, location, wind speed) into descriptive natural language and utilized multi-modal contrastive learning to learn from the supervision information in the form of text. This work explores the feasibility of using contrastive learning for multi-feature learning.

\section{Methods}

\begin{table*}[t]
\normalsize
  \centering
  \caption{Batch size and feature dimensions.}
  \resizebox{\linewidth}{!}{
  \begin{threeparttable}
  \setlength\extrarowheight{2pt}
    \begin{tabular}{lllll}
    \toprule
    Methods & Features & Shipsear (batch size/dim)  & DeepShip (batch size/dim) & DTIL (batch size/dim)\\
    \midrule
    Baseline & STFT& 32/(1199,1318)&64/(1199,400)&128(1199,100) \\
      & Mel& 32/(1199,300)&64/(1199,300)&128/(1199,300)\\
      & CQT& 32/(899,340)&64/(899,289)&128(899,229)\\
    Model ensemble & Mel, CQT& 32/(1199,300)+(899,340)&64/(1199,300)+(899,289)&128/(1199,300)+(899,229)\\
    Contrastive learning& Mel, CQT& 32/(1199,300)+(899,340)&64/(1199,300)+(899,289)&128/(1199,300)+(899,229)\\

    \bottomrule
    \end{tabular}
    \end{threeparttable}}
  \label{tab_para}%
  \vspace{-3mm}
\end{table*}%
The CAM in Fig.~\ref{fig1} displays the focus of the recognition model on the input STFT spectrograms. To enhance the low-frequency line spectrum information and high-frequency periodic modulation information, we extract two spectrogram-based features with different emphases as the inputs of contrastive learning. The complete processing pipeline is illustrated in Fig.~\ref{fig2} and will be detailed introduced as follows:

At the beginning, we collect the underwater acoustic signals (e.g., ship-radiated noise) detected by passive sonar arrays. To enhance the signals in target directions and suppress interference and noise from other directions, array signal processing techniques, such as beamforming, are applied during the preprocessing stage. Then, we get the enhanced single-channel signals for subsequent feature extraction and recognition.

The next step is feature extraction. Given a batch of $N$ samples, we mark the $i$th audio sample as ($x_i,y_i$) and apply the short-time Fourier transform to compute the corresponding FFT spectrum.  We first apply Mel filter banks to the spectrum to implement Mel filtering. As depicted in Equation (1), the Mel filter bank is a set of triangular bandpass filters that are spaced according to the non-linear Mel scale. The Mel filter banks are denser at low frequencies, resulting in higher frequency resolution. Next, the filtered spectrum is transformed using a logarithmic scale to get the Mel spectrograms $Mel_i$.

\begin{equation}
    Mel(f)=2595\times log(1+\frac{f}{700})
\end{equation}

In the second step, we obtain the CQT spectrogram by conducting the constant Q transform. We convolve the FFT spectrum of each frame with the CQT kernel, which is a bank of bandpass filters that are logarithmically spaced in frequency. Denote the maximum or minimum frequency to be processed as $f_{max}$,$f_{min}$. The $k-$th frequency component $f_k$ can be formalized as:

\begin{equation}
    f_k=2^{k/b} f_{min} \quad f_{min}\leq f_k\leq f_{max}
\end{equation}

where b is the octave resolution. The magnitude of the filtered spectrum is used to represent the CQT spectrogram $CQT_i$. The CQT spectrogram has high temporal resolution at high frequencies.

After extracting the Mel spectrograms and CQT spectrograms to focus on the low-frequency and high-frequency information, we employ two corresponding encoders $F1(\cdot)$, $F2(\cdot)$ to learn from the two features simultaneously. Fig.~\ref{fig2} describes the complete training pipeline in detail. we feed a batch of $Mel_i$ and $CQT_i$ into the corresponding encoders $F1(\cdot)$ and $F2(\cdot)$, and obtain two embedding vectors $E_1=F1(Mel_i),E_2=F2(CQT_i)$ (the dimension of embedding vectors is $N\times512$). Then, we apply the cosine similarity-based contrastive learning by computing the cosine similarity matrix $M$ (the dimension of $M$ is $N\times N$) between the two embeddings:

\begin{equation}
    M = \frac{E_{1}\cdot E_{2}^\top}{\Vert E_{1} \Vert_2 \times \Vert E_{2} \Vert_2} 
    %= 
    %\sum_{i=1}^N\frac{E_{1_i}}{\Vert E_{1} %\Vert_2}\frac{E_{2_i}^\top}{\Vert E_{2} \Vert_2}
\end{equation}

As depicted in Equation (3), we apply a normalization operation to unify the range of embedding values during calculating the similarity matrix. Our optimization goal is to maximize the similarity between the two normalized embeddings from the same sample $E1_i$ (dim=1$\times$512) and $E2_i$ (dim=1$\times$512), while minimizing the similarity between two embeddings from different samples $E1_i$ and $E2_j (j\neq i)$. From the perspective of the similarity matrix, the optimization goal is to maximize the elements on the diagonal and minimize the other elements in the matrix. Thus, we define the loss function as $\mathcal{L}_{ICL}$ by computing the cross-entropy loss between the similarity matrix $M$ and an identity matrix $I_N$.

After that, we combine the embeddings $E=E_1+E_2$ and feed $E$ into a task-dependent linear fully connected layer $Fc(\cdot)$. We could get $\mathcal{L}_{CE}$ by computing the cross-entropy loss between the output logits $logits=Fc(E)$ and the ground truth $y$. The final loss could be depicted by a simple weighted addition:

\begin{equation}
\begin{aligned}
    \mathcal{L} 
    &= \mathcal{L}_{CE}+\alpha \mathcal{L}_{ICL} \\
    &=CE(logits,y)+\alpha CE(M,I_N)
\end{aligned}
\end{equation}

where $\alpha$ is an adjustable coefficient used to control the influence of contrastive constraints on the overall model learning. Regarding the selection of the value of $\alpha$, we conduct relevant experiments in Section V.

\section{Experiment Setup}

\subsection{Datasets}
In this work, we use three underwater ship-radiated noise datasets of different scales. The detailed information is listed in the following paragraphs:

1. Shipsear\cite{b6} is an open-source database of underwater recordings of ship and boat sounds. The database is composed of 90 records representing sounds from 11 vessel types. It consists of nearly 3 hours of recordings. Considering that too little data is difficult to cut into the form of ``train, validation, test'', we select a subset of 9 categories (dredger, fish boat, motorboat, mussel boat,natural noise, ocean liner, passenger ship, ro-ro ship, sailboat) in Shipsear for the recognition task.

2. DeepShip\cite{b5} is an open-source underwater acoustic benchmark dataset, which consists of 47.07 hours of real-world underwater recordings of 265 different ships belonging to four categories (cargo, passenger ship, tanker, and tug).

3. Our private dataset\cite{b22}: data collected from Thousand Island Lake (DTIL) is a dataset collected from Thousand Island Lake, which contains multiple sources of interference. The dataset contains 330 minutes of the speedboats and 285 minutes of the experimental vessels.

\subsection{Baseline Methods}
For the recognition tasks, we adopt the convolutional neural network with residual structure - ResNet\cite{b9} as our baseline model, which shows superior performance on spectrogram-based recognition\cite{b1}\cite{b7}\cite{b23}. Likewise, our ICL also takes ResNet as the backbone of encoders to ensure fair comparisons. For the ICL encoders, we remove the last fully-connected layer of ResNet to obtain high-dimensional embeddings. 

Furthermore, considering that our contrastive learning employs multiple acoustic features, we need to set a baseline using multiple acoustic features for a fair comparison. In this work, we take the decision-level model ensemble as our baseline method. This is a simple technique that ensembles predictions of multiple models. When two models predict the same category, the ensembled model makes the same prediction; When the two predictions vary, the prediction with higher confidence is taken as the final prediction.

\subsection{Parameters Setup}
Our parameter setup for batch sizes and the feature dimensions is illustrated in Table~\ref{tab_para}.  Due to the differences in sampling rates and effective frequency bands on the three datasets, the corresponding feature dimensions vary greatly. Based on the amount of data and feature dimensions, we adopt different batch sizes for three datasets. When the feature dimension is large, we need to set a small batch size to ensure that the device does not run out of memory.

In this work, each signal is cut into 30-second segments with a 15-second overlap. To avoid information leakage, we make sure that segments in the training set and the test set must not belong to the same audio track. During framing, this work sets the frame length as 50ms and the frame shift as 25ms. Besides, we set the number of Mel filter banks to 300, and the octave resolution of CQT to 36 as default. During training, we use the AdamW\cite{b24} optimizer with weight decay. The learning rate is set to 5e-4 and the weight decay is set to 1e-5 for all experiments. All models are trained for 200 epochs on four A40 GPUs.

\section{Experiments and Results}
To demonstrate the effectiveness of our ICL, we conduct detailed comparative experiments on three underwater databases. For all experiments, we uniformly take accuracy as the evaluation metric.

\subsection{Main Results}
\begin{table}[t]
\normalsize
  \centering
  \caption{Main results on three underwater acoustic databases.}
  \resizebox{\linewidth}{!}{
  \begin{threeparttable}
  \setlength\extrarowheight{2pt}
    \begin{tabular}{llccc}
    \toprule
    Methods & Features & Shipsear & DeepShip & DTIL\\
    \midrule
    Baseline & STFT& 75.24 & 74.68& 95.93\\
      & Mel& 77.14 & 74.85& 95.48\\
      & CQT& 73.33 & 77.82& 96.48\\
    Model ensemble & Mel, CQT& 78.67 & 78.28& 97.18\\
    Contrastive learning& Mel, CQT& \textbf{85.34} & \textbf{80.02}& \textbf{98.87}\\

    \bottomrule
    \end{tabular}
    \end{threeparttable}}
  \label{tab1}%
  \vspace{-3mm}
\end{table}%

In Table~\ref{tab1}, we present the results of our experiments on three datasets using various features and methods. First, we observe that models based on Mel and CQT spectrograms have a certain degree of improvement compared to those based on the original STFT spectrogram on three datasets, indicating that the application of filters is beneficial for recognition.

Furthermore, we perform model ensemble on the Mel and CQT spectrogram-based models and find that the multi-feature-based recognition models can outperform those based on a single feature. However, the gains achieved through model ensemble are not significant. By contrast, the application of contrastive learning (ICL) consistently leads to a promising performance boost across all three datasets. This suggests that taking advantage of  complementary features through contrastive learning is more effective than learning from multiple features independently.

%It leads recognition models to  learn from a more comprehensive and non-biased perspective. Experimental results demonstrate that ICL, inspired by the class activation mapping of spectrograms, can fully exploit the potential of learning from multiple features.
%, which conforms to the preference for network learning (see Fig.~\ref{fig1}).
% can force two encoders with different concerns to constrain each other,

\begin{figure}
    \centering
    \subfigure[Confusion matrix heat maps for single-feature models on Shipsear. On the left is the baseline model with Mel spectrogram, and on the right is the baseline model with CQT spectrogram.]{
        \begin{minipage}[b]{0.45\textwidth}
        \includegraphics[width=\linewidth]{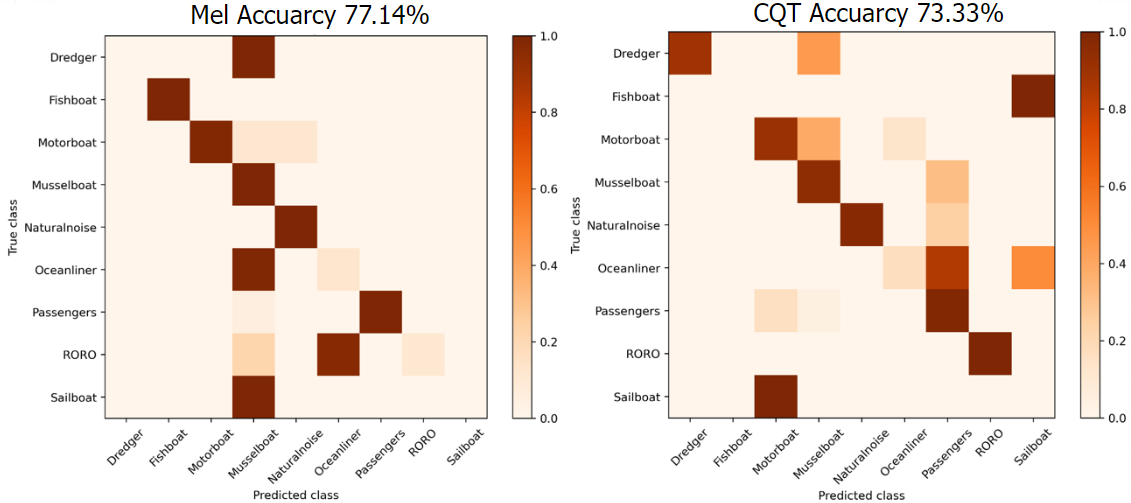}
        \end{minipage}}

    \subfigure[Confusion matrix heat maps for multi-feature models on Shipsear. On the left is the ensembled model with Mel and CQT spectrogram, and on the right is the model guided by ICL with Mel and CQT spectrogram.]{
        \begin{minipage}[b]{0.45\textwidth}
        \includegraphics[width=\linewidth]{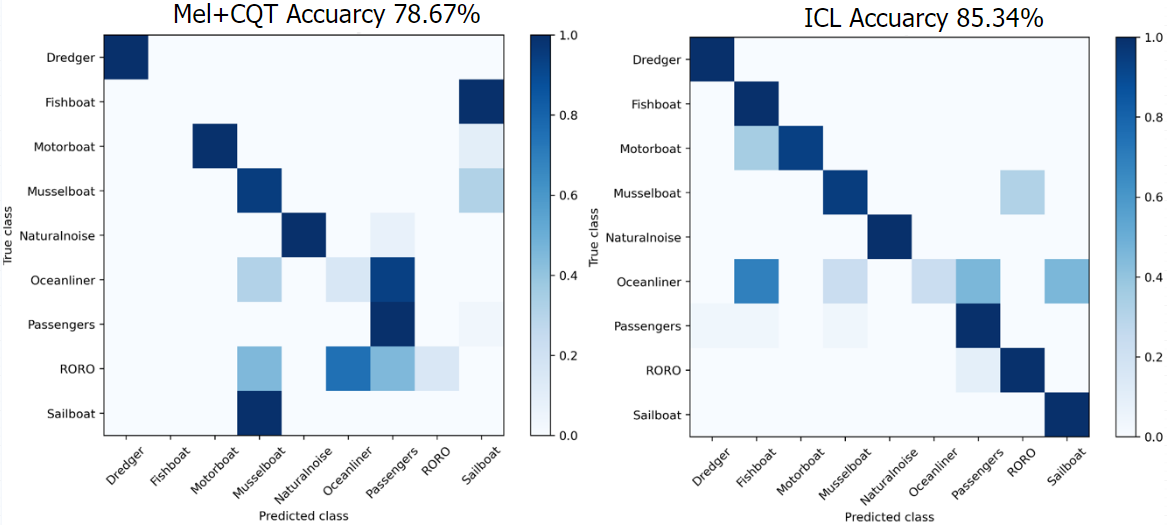}
        \end{minipage}}
    \caption{Confusion matrix heat maps.}
    \label{fig3}
    \vspace{-2px}
\end{figure}

Besides, we find that ICL can bring a significant boost to the data-scarce Shipsear, while bringing less promotion to the data-sufficient DeepShip. On data-scarce Shipsear, our contrastive learning technique (85.37\%) can bring an improvement of 8.62\% on the raw single-feature model (76.72\%). According to our analysis, in cases where data is scarce, models are prone to overfitting and are likely to make incorrect predictions on unseen data. Model ensembles are powerless against the overfitting, but ICL's regularization effect can constrain the encoders with an extra loss term during training and mitigate the negative impact of overfitting. When there is sufficient data (such as in DeepShip), overfitting does not significantly impact performance and ICL's regularization ability is not displayed, resulting in less improvement.

\subsection{Visualization}

\begin{table}[t]
\normalsize
  \centering
  \caption{Feature selection for contrastive learning.}
  \resizebox{0.99\linewidth}{!}{
  \begin{threeparttable}
  \setlength\extrarowheight{2pt}
    \begin{tabular}{llccc}
    \toprule
    Methods & Features & Shipsear & DeepShip & DTIL\\
    \midrule
    Contrastive learning & STFT, CQT& 84.48 & 78.72& 97.61\\
    - & STFT, Mel& 83.62 & 77.11& 98.02\\
    -& Mel, CQT& \textbf{85.34} & \textbf{80.02}& \textbf{98.87}\\

    \bottomrule
    \end{tabular}
    \end{threeparttable}}
  \label{tab2}%
  \vspace{-3mm}
\end{table}%

To further demonstrate the superiority of our ICL technology, we plot the confusion matrix heat maps for four models: 1. baseline model with Mel spectrogram; 2. baseline model with CQT spectrogram; 3. ensembled model with Mel and CQT spectrogram; 4. model guided by ICL with Mel and CQT spectrogram. The confusion matrix compares the true labels (y-axis) to the predicted labels (x-axis) and summarizes the results in a clear and concise manner. The confusion matrix heat map (see Fig.~\ref{fig3}) is the graphical representation of the confusion matrix, where the values in the matrix are represented as colors. The color scale used in a heatmap ranges from light (representing low values) to dark (representing high values).

As illustrated in Fig.~\ref{fig3}(a), single-feature models often make wrong predictions. The model based on the Mel spectrogram can not correctly recognize dredgers, ocean liners, ro-ro ships, and sailboats, while the model based on the CQT spectrogram is not good at recognizing fish boats, ocean liners, and sailboats. Fig.~\ref{fig3}(b) reveals that the ensembled model can indeed bring a little performance boost, but the prediction of certain categories may be misled by one of the biased models. For example, the model based on the Mel spectrogram can recognize fish boats correctly, while the ensembled model suffers from the bias of the CQT-based model (recognize fish boats as sailboats) and fails to correctly recognize fish boats. It shows that model ensembles are not good at compensating for overfitting models, and instead run the risk of being misguided. For our ICL-based model, ICL constrains the model at the training stage and has a regularization effect. The two encoders may reduce overfitting under the constraints so that the overall model can have better recognition ability. As shown in Fig.~\ref{fig3}(b), the ICL-based model can correctly recognize most categories even for categories that are not correctly recognized by all single-feature models (e.g., sailboats).

\subsection{Feature Selection}
As presented in Table~\ref{tab1}, the Mel spectrogram with higher frequency resolution in the low-frequency component is suitable for Shipsear, and the CQT spectrogram with higher time resolution in the high-frequency component performs better on DeepShip and DTIL. In most instances, the two features with different emphases can both achieve better results than the original STFT spectrogram. The original spectrogram has no emphases, and its information redundancy may result in a slightly inferior performance.

Besides, we conduct further investigation into the impact of features used in contrastive learning. As illustrated in Table~\ref{tab2}, we observe that applying both Mel and CQT spectrogram is the most effective strategy. Because the information they focus on is complementary, which is beneficial to help the recognition system learn from more comprehensive aspects. The results further prove that it is a suitable and effective choice for us to choose Mel and CQT spectrogram as the input of contrastive learning according to the information contained in CAMs.

\subsection{Selection of the Adjustable Coefficient}
According to Equation (4), $\alpha$ is an adjustable coefficient used to control the influence of contrastive constraints on the final target loss. The value of $\alpha$ has an obvious influence on the model training. If the value of $\alpha$ is too small, the potential of contrastive learning cannot be fully utilized; if the value of $\alpha$ is too large, the model may pay too much attention to the relationship between the two features, rather than focusing on the target recognition task. Therefore, it is very necessary to choose a moderate value of $\alpha$.

\begin{table}[t]
\normalsize
  \centering
  \caption{Experiments on the adjustable coefficient on Shipsear.}
  \resizebox{\linewidth}{!}{
  \begin{threeparttable}
  \setlength\extrarowheight{2pt}
    \begin{tabular}{cc}
    \toprule
    Value of the adjustable coefficient & Accuracy on Shipsear\\
    \midrule
    0.2 & 83.62\\
    0.5 & \textbf{85.34} \\
    1 & 84.48\\
    2 & 82.76\\
    \bottomrule
    \end{tabular}
    \end{threeparttable}}
  \label{tab3}%
  \vspace{-3mm}
\end{table}%

Table~\ref{tab3} shows that $\alpha=0.5$ is a good choice. This moderate value can give full play to the role of contrastive learning, while balancing contrastive learning and the target recognition task.

\subsection{Training Time Cost}

\begin{table}[t]
\Large
  \centering
  \caption{Training time cost.}
  \resizebox{\linewidth}{!}{
  \begin{threeparttable}
  \setlength\extrarowheight{2pt}
    \begin{tabular}{lllll}
    \toprule
    Methods & Features & Shipsear (mins)  & DeepShip (mins) & DTIL (mins)\\
    \midrule
    Baseline & STFT& 126.5&302.2&44.0 \\
      & Mel& 36.4&370.6&128.2\\
      & CQT& 28.5&254.4&68.8\\
    Model ensemble & Mel, CQT& 64.9&625.0&197.0\\
    Contrastive learning& Mel, CQT& 57.1&468.3&167.6\\

    \bottomrule
    \end{tabular}
    \end{threeparttable}}
  \label{tab_cost}%
  \vspace{-3mm}
\end{table}%
In this subsection, we analyze the training time cost of our experiments. As shown in Table~\ref{tab_cost}, the training time varies due to the different batch sizes, feature dimensions (see Table~\ref{tab_para}), data quantity, and algorithm efficiency. It can be inferred from the results in the table that the efficiency of ICL is satisfactory. The model based on ICL has a similar number of parameters as the ensembled model, but its training time consumption is significantly lower. The two encoders used in contrastive learning can be trained in parallel, which greatly improves the training efficiency.

\section{Conclusion}
In this work, we first apply the class activation mapping to highlight and visualize the critical regions in the spectrogram. The CAMs reveal that the model learning is guided by the low-frequency line spectrum information and high-frequency periodic modulation information. Then we take it as a starting point and propose interpretable contrastive learning (ICL) based on the Mel and CQT spectrogram. ICL can force two spectrograms with different emphases to constrain each other, thus achieving the regularization effect and improving the generalization performance. Our experiments and related visualization analysis show that ICL can achieve promising results on underwater target recognition tasks, especially when data is scarce.

As a deficiency, this work only adopts simple contrastive learning based on cosine similarity. In the future, we plan to try more advanced contrastive learning strategies to further improve the recognition ability of deep learning models in real scenarios.

\section{Acknowledgements}
This research is supported by the IOA Frontier Exploration Project (No. ZYTS202001) and Youth Innovation Promotion Association CAS.

%\vspace{12pt}
%\color{red}
%IEEE conference templates contain guidance text for composing and formatting conference papers. Please ensure that all template text is removed from your conference paper prior to submission to the conference. Failure to remove the template text from your paper may result in your paper not being published.

\end{document}